\documentclass[10pt]{article}
\usepackage{amsfonts}
\usepackage{amssymb}
\usepackage{amsthm}

\usepackage{graphicx}

\newcommand{\sss}{\setcounter{equation}{0}}
\newtheorem{theorem}{THEOREM}[section]

\newtheorem{lemma}[theorem]{LEMMA}

\newtheorem{Ansatz}[theorem]{The Aharonov-Bohm Ansatz}

\newcommand{\ere}{ {\mathbb R}}

\def\beq{\begin{equation}}
\def\ene{\end{equation}}
\def \ds {\displaystyle}
\newcommand{\bull}{\hfill $\Box$}
\def\v{\mathbf v}

\def\hv{\hat{\mathbf v}}

\def\mo{\mathbf p}

\def\tvf{\tilde{\varphi}}

\def\+out{x^{\rm out}}
\def\Vab{V_{AB}}
\def\tv{\tilde{v}}
\def\tbv{\tilde{\mathbf v}}
\begin{document}
\baselineskip=20 pt
\parskip 6 pt

\title{The Electric Aharonov-Bohm Effect
\thanks{ PACS Classification (2008): 03.65.Nk, 03.65.Ca, 03.65.Db, 03.65.Ta.  Mathematics Subject Classification(2000): 81U40, 35P25,
35Q40, 35R30.}
\thanks{ Research partially supported by
 CONACYT under Project CB-2008-01-99100.}}
\author{ Ricardo Weder\thanks {Fellow, Sistema Nacional de Investigadores.
Electronic mail: weder@servidor.unam.mx  }\\
Departamento de M\'etodos
 Matem\'aticos  y Num\'ericos.\\
 Instituto de Investigaciones en Matem\'aticas Aplicadas y en
 Sistemas. \\
 Universidad Nacional Aut\'onoma de M\'exico.\\
  Apartado Postal 20-726,
M\'exico DF 01000, M\'exico.}

\date{}
\maketitle


\vspace{.5cm}
 \centerline{{\bf Abstract}}
\bigskip
\noindent The seminal paper of Aharonov and Bohm [Significance of
electromagnetic potentials in the quantum theory, Phys. Rev. {\bf
115} (1959) 485-491 ] is at the origin of a very extensive
literature in some of the more fundamental issues in physics. They
claimed that  electromagnetic fields can  {\it  act at a distance}
on charged particles even if they are identically zero in the region
of space where the particles propagate, that  the fundamental
electromagnetic quantities in quantum physics are not only the
electromagnetic fields  but also the circulations of the
electromagnetic potentials; what gives  them  a real physical
significance. They proposed two experiments to verify their
theoretical conclusions. The magnetic Aharonov-Bohm  effect, where
an electron is influenced by a magnetic field  that is zero in the
region of space   accessible to the electron, and the electric
Aharonov-Bohm  effect where an electron is affected by a
time-dependent  electric potential that is constant in the
region where the electron is propagating, i.e., such that the
electric field vanishes along its trajectory. The Aharonov-Bohm
effects imply such a strong departure from the physical intuition
coming from classical physics that it is no wonder that they remain
a highly controversial issue after more than fifty years, on spite
of the fact that they are discussed in most of the text books in
quantum mechanics. The magnetic case has been extensively studied.
The experimental  issues were settled by the remarkable experiments
of Tonomura et al. [Observation of Aharonov-Bohm effect by electron
holography, Phys. Rev. Lett. {\bf 48} (1982) 1443-1446 , Evidence
for Aharonov-Bohm effect with magnetic field completely shielded
from electron wave, Phys. Rev. Lett. {\bf 56} (1986) 792-795] with
toroidal magnets, that gave a strong  evidence of the existence of
the effect, and by the recent experiment of Caprez et al.
[Macroscopic test of the Aharonov-Bohm effect, Phys. Rev. Lett. {\bf
99}  (2007) 210401] that shows that the results of the Tonomura et
al. experiments can not be explained by the action of a force. The
theoretical issues were settled Ballesteros and Weder [High-velocity
estimates for the scattering operator and Aharonov-Bohm effect in
three dimensions, Comm. Math. Phys. {\bf 285} (2009) 345-398, The
Aharonov-Bohm effect and Tonomura et al. experiments: Rigorous
results, J. Math. Phys. {\bf 50} (2009) 122108, Aharonov-Bohm Effect
and High-Velocity Estimates of Solutions to the Schr\"odinger
Equation,  Commun. Math. Phys. {\bf 303} (2011) 175-211] who rigorously  proved  that
quantum mechanics predicts the experimental results of Tonomura et
al. and of Caprez et al.. The electric Aharonov-Bohm effect has been
much less studied. Actually, its existence, that has not been
confirmed experimentally, is a very controversial issue. In their
1959 paper Aharonov and Bohm proposed an Ansatz for the solution to
the Schr\"odinger equation in regions  where there is a
time-dependent  electric potential that is constant in space. It
consists in multiplying the free evolution by a phase given by the
integral in time of the potential. The validity of this Ansatz
predicts interference fringes between parts of a coherent electron
beam that are subjected to different potentials. In this paper we prove that the exact solution to the
Schr\"odinger equation is given by the Aharonov-Bohm Ansatz up to an
error bound in norm that is uniform in time and that decays as a
constant divided by the velocity. Our results give, for the first time, a rigorous proof
that quantum mechanics  predicts the existence of the electric
Aharonov-Bohm effect, under conditions that we provide. We hope that
our results will stimulate the experimental research on the
electric Aharonov-Bohm effect.

\section{Introduction}\sss
In classical electrodynamics the evolution of a charged particle in the
presence of an electric field, $E$, is given by Newton's equation
with the force $F= q E$, where $q$ is the charge of the particle. If
a particle  propagates in a region were the electric field is zero
the force is zero and the trajectory  is a straight line. The fundamental physical quantity is the
electric field and, of course, also the magnetic field if there is
one. Let $\mathbf V$ be an electric potential such that $ E=-\nabla {\mathbf V}$. Newton's
equation implies that the trajectory of a classical charged particle
is not affected by an electric potential that is constant in the
region of propagation, since in this case $F= q E= -q \nabla {\mathbf V} =0$.

In quantum physics the situation is quite different. Quantum
mechanics is a Hamiltonian theory were the dynamics of a charged
particle in the presence of an electric  field is governed by
Schr\"odinger's equation  that can not be formulated directly in
terms of the electric  field. The introduction of an electric
potential is required to define the Hamiltonian. Aharonov and Bohm
observed \cite{ab} that this raises the possibility that a
(time-dependent) electric potential could act on a charged particle
even if it is constant in the region of space where the particle
propagates, and they  proposed an experiment to
verify their theoretical prediction. See Figure 1. They advised to
split a coherent electron beam into two parts and to let each one
enter a long cylindrical metal tube. When each beam is well inside
its tube, electric potentials are applied, in such a way   that at
any given time they are  constant in the part of the tubes  where
each beam is propagating. The potentials are set to zero well before
the beams leave the tubes. Finally, after the beams leave the tubes
they are combined to interfere coherently. They claimed that as the
potentials are constant in the region of the tubes where the beams
propagate, the tubes  act as a Faraday cage, and each beam
picks up a phase given by the integral in time of its potential. If
the potentials are different, so are the phases, and an
interference fringe  should be produced. This interference fringe is a
purely quantum mechanical phenomenon, because on this experiment the
beams are in a time-varying potential without ever being in an
electric field, since the field does not penetrates far from the
edges of the tubes, and it is only non-zero at times when the beams
are well inside the tubes, far from the edges, what means that
classically no force acts on the electron beams. This is the
electric Aharonov-Bohm effect. In the same paper \cite{ab} they also
proposed an experiment where a coherent electron beam is split into
two beams and each one is allowed to pass, respectively, to the left and to the
right of a magnetic field that is zero along the path of each beam.
When the beams are behind the magnetic field they are combined  to
interfere. They predicted that an interference fringe will be
observed that it is due to the {\it action at a distance} of the
magnetic field and that it will depend on the circulation of the
magnetic potential, what gives magnetic potentials a physical
significance. This, of course, is impossible in classical physics.
This is the magnetic Aharonov-Bohm effect. Note, however, that the
existence of these interference fringes was previously  predicted by
Franz \cite{f}.

The experimental verification of the Aharonov-Bohm effects
constitutes a test of the validity of the theory of quantum
mechanics itself. For a review of the literature up to 1989 see
\cite{op} and \cite{pt}. In particular, in \cite{pt} there is a
detailed discussion of the large controversy -involving over three
hundred papers up to 1989- concerning the existence of the
Aharonov-Bohm effect. For a recent update of this controversy see
\cite{bt,to,tn}.

As mentioned in the abstract, the magnetic case has been extensively studied, but even the existence of the electric Aharonov-Bohm effect is questioned.  Note that in the experiment \cite{mp}
a steady-state version of the electric Aharonov-Bohm effect was
tested and the expected phase shift was observed. However, as it was pointed out in \cite{bt},
in the steady-state electric Aharonov-Bohm effect the electrons are
subjected to a force and, for this reason, it is not considered to
be a verification of the electric Aharonov-Bohm effect, where no
forces act on the electrons. 

As pointed out above,  above, Aharonov and Bohm \cite{ab} proposed an Ansatz
for the solution to the Schr\"odinger equation  in  regions where
there is a time-dependent electric potential that  is constant
 in space. It consists of multiplying the free evolution by a phase given by the integral in
time of the potential. As the Aharonov-Bohm  Ansatz predicts an
interference fringe between the different parts of a coherent beam
that are subjected to different potentials, the issue of the
existence of the electric Aharonov-Bohm effect  can  be summarized
in a single mathematical question: is the Aharonov-Bohm Ansatz a
good approximation to the exact solution to the Schr\"odinger
equation, under the conditions of the experiment proposed by
Aharonov and Bohm. This is the question that we address in this
paper.

Let us consider the case of one electron beam and one tube, $K$, or,
equivalently, the part of the electron beam that travels inside one
of the tubes, after splitting the original beam into two. For the
Aharonov-Bohm Ansatz to be valid it is necessary that, to a good
approximation, the electron does not interact with  $K$, because if
it hits  $K$ it will be reflected and the solution can not be the
free evolution multiplied by a phase. This is true no matter how big
the velocity is. For this reason   we consider a general class of
incoming asymptotic states with the property that under the free
classical evolution they do not hit $K$. The intuition is that for
high velocity the exact quantum mechanical evolution is close to the
free quantum mechanical evolution and that as the free quantum
mechanical evolution is concentrated on the classical trajectories,
we can expect that, in the leading order for high velocity, we do
not see the influence of $K$ and that only the influence of electric
potential  inside $K$ shows up in the form of a phase, as predicted
by the Aharonov-Bohm Ansatz.

We prove in this paper that the exact
solution to the Schr\"odinger equation  is given by the
Aharonov-Bohm Ansatz, up to an error bound in norm that is uniform
in time and that decays as a constant divided by the velocity  $v$.
In our bound the direction of the velocity is kept fixed, along the
axis of the tube, as it absolute value goes to infinite.

We study this problem in $\ere^n, n \geq 2,$ because the proofs are
the same for all $n \geq 2$, but of course, the physical case is
$n=3$.

Let us denote  $\mo:=-i\nabla$. The Schr\"odinger equation
for an electron in $\Lambda:= \ere^n \setminus K$,  with electric potential ${\mathbf V}(t,x)$   is given by
\beq
i\hbar
\frac{\partial}{\partial t} \phi = \frac{1}{2 M} {\mathbf P}^2 \phi+ q \,\mathbf{V}(t,x)\phi,
\label{1.0}
\ene
where $\hbar$ is Planck's constant, ${\mathbf P}:=\hbar \mo$ is the
momentum operator, and $M$ and $q$ are,
respectively, the mass and the charge of the electron.

Suppose that  $K$ is centered at the origin, $x=0$, and that its
axis is along the vertical coordinate, $x_n$. Furthermore, assume
that the velocity of the electron is along $x_n$ and that at time
zero it is localized well inside $K$, in a neighborhood of  the
origin. Let $\mathbf V_{AB}$  be the electric potential in the experiment
proposed by Aharonov and Bohm. $\mathbf{V}_{AB}$ is zero before the electron
enters  $K$, then it grows in time when the electron is well inside
$K$, and finally if falls back to zero before the electron comes
near the other edge of $K$. Note that the electron is inside the
tube during a time interval, around zero, of the order $1/v$. Hence,
$\mathbf V_{AB}$ is different from zero only during a time interval of the
order $1/v$. Since as $v$ increases the time that $\mathbf V_{AB}$ acts on the
electron decreases as $1/v$, in order that its effect does not
disappears for large $v$ it is necessary that the strength of $\mathbf V_{AB}$
increases with the velocity $v$. For this reason we take $\mathbf V_{AB}$ as
follows,

\beq \label{1.1}
 \mathbf V_{AB}(vt,x):= v \,\mathbf Q(v t,x).
 \ene
 We denote by $K_0$ the hole of $K$ . Let $B_R$ denote  the open ball of center zero and radius $R$.
We assume that  for some  $L_1 >L_0 >0$,  such that $B_{L_1} \subset K_0$, we have that   $ \mathbf Q(z,x)=0, |z|\geq
L_0$ and that  for  $|z| <  L_0,\,\, \mathbf Q(z,x)= \mathbf Q_0(z)$   for $ x \in B_{L_1}$, where $\mathbf Q_0(z)$ is a continuously differentiable  function that vanishes for $|z| \geq L_0$. Note that $z=vt$ is the  distance along the classical trajectory of an electron that propagates with velocity $\v$.

Since  high-velocity estimates of solutions to  Schr\"odinger
equations are of independent interest, we consider a situation that
goes beyond the electric Aharonov-Bohm effect and assume that the
electric potential is of the form,

\beq
\mathbf V(x,t):= \mathbf V_{AB}(vt,x)+ \mathbf V_0(t,x).
\label{1.2}
\ene
where $\mathbf V_0$ is a  potential that is  uniformly bounded  on the velocity. As we prove below $\mathbf V_0$  gives  no contribution to the leading order for high velocity, and then, on this regime, it plays no role in the electric Aharonov-Bohm effect.

The free Hamiltonian $\mathbf H_0$ is given by
$$
\mathbf H_0:=  \frac{1}{2M}\,\mathbf P^2.
$$

The incoming free electron beam with velocity $\v$ is given by
$$
\psi_{\v,0}= e^{-i \frac{t}{\hbar}\,\mathbf H_0}\, \varphi_\v,
$$
where
$$
\varphi_\v= e^{i  \frac{M}{\hbar}\v\cdot x}\, \varphi.
$$
We designate by  $\Lambda$  the complement  of the tube:  $\Lambda:= \ere^n \setminus K$.
For any $\v \neq 0$ we denote,
$$
\Lambda_{\hv}:= \{x \in \Lambda:
x+\tau \hv \in \Lambda,\, \forall \tau \in \ere\},
$$
where $\hv:= \v /|\v|$. We show in Subsection 3.1  that the incoming free electron beams $\psi_{\v,0}$ with support $\varphi \subset \Lambda_\v$ have negligible interaction with the cylinder $K$ for large velocities if the wave packet spreading is neglected. In fact, it is only for this type of incoming electron beams that we can expect that the Aharonov-Bohm Ansatz  is a good approximation to the exact solution for large velocities.
 
 We denote,
$$
F_-(t):= v \int_{-\infty}^{t}\, \frac{q}{\hbar} \mathbf Q_0(v s)\, ds.
$$
The Aharonov-Bohm Ansatz   is given by,
$$
\psi_{AB,\v}(t,x) := e^{-i \, v \int_{-\infty}^{t}\, \frac{q}{\hbar} \mathbf Q_0(v s)\, ds}\, e^{-i \frac{t}{\hbar} \,\mathbf H_0}\, \varphi_{\v}.
$$
 The unique solution to the Schr\"odinger equation  (\ref{1.0})   that behaves as the free incoming
electron beam, $\psi_{\v,0}$, as $ t \rightarrow  -\infty$ is given by

$$
\psi_\v:= U(t,0)\, W_{-} \,\varphi_{\v},
$$
where   $U(t,0)$ is the propagator for  (\ref{1.0})  and $W_{-} $ is a wave operator. See  Subsection 2.3 and equations (\ref{2.26}, \ref{3.3},  \ref{3.4}).
 
By  Theorem \ref{theor-3.1} in Subsection 3.3, for any $\v \in \ere^n \setminus 0$ such that $B_{L_1} \subset \Lambda_{\v}$ and for any $ 0 < R <  L_1-L_0$ there is a constant $C$ such that,
$$
\left\|U(t,0)\,
\,\psi_{\v} - \psi_{AB,\v}\right\| \leq \,C \, \|\varphi\|_{ \mathcal H_2(\ere^n)}\, \,{\mathcal E}(v),
$$
for all  $\varphi$ in the Sobolev space  $\mathcal H_2(\ere^n)$ with support contained in $B_R$, and 
where the error ${\mathcal E}(v)$ is given by,
\beq
{\mathcal E}(v):= \left\{
 \begin{array}{c} \ds \frac{1}{v^\rho}, \quad 0 < \rho < 1,\\\\
\ds \frac{|\ln v|}{v}, \quad \rho=1,\\\\
 \ds\frac{1}{v}, \quad \rho > 1,
 \end{array}
 \right.
\label{3.11b}
\ene
for $ v >0$ and where $\rho$  gives the decay rate of $\mathbf V_0$ as $|x| \rightarrow \infty$.  See equation (\ref{2.6}). 

Note that $\mathcal E= 1/v$ if $V_0$  decays as a short-range potential at infinite. Actually, for the purpose of the Aharonov-Bohm effect we can take
 $V_0=0$. We give a precise definition of $K$ in Subsection 2.1 and in Subsection 2.2 we state our conditions in the electric potential $\mathbf V$.

Let us consider the experiment proposed by Aharonov and Bohm \cite{ab} in three dimensions with one cylinder with axis along the vertical coordinate
$x_3$ and let us take $\v$ directed along $x_3$. We consider an incident coherent electron beam that we  split into two parts. One travels inside the
tube where it is influenced by the Aharonov-Bohm potential and the other, that is the reference beam, travels outside the tube where the Aharonov-Bohm
potential is zero. Finally, both parts are brought together behind the tube and are allowed to interfere. We can equivalently consider that both beams
 travel inside the tube, one with the Aharonov-Bohm potential and the other without it.
For high velocity $\psi_\v$ is well approximated by
$\psi_{AB,\v}$. Furthermore, behind the tube,
$$
\psi_{AB,\v}=\, e^{-i \frac{q}{\hbar} \mathbf \Phi}\, e^{-i \frac{t}{\hbar} \,\mathbf H_0}\, \varphi_{\v}, \, \hbox{\rm for}\, t \geq L_0/v,\,
\hbox{\rm or}\, z=vt \geq L_0
$$
where,
$$
\mathbf \Phi:=\int_{-L_0}^{L_0}\, \mathbf Q_0(z)\, dz.
$$
The reference beam is given by,
$$
 e^{-i \frac{t}{\hbar} \,\mathbf H_0}\, \varphi_{\v}.
$$
We see that the beam that travels inside the tube with the Aharonov-Bohm potential, and the reference beam show precisely the difference in phase
 predicted by Aharonov and Bohm \cite{ab}. Our results prove, for the first time, that quantum mechanics rigorously predicts the existence of the
 electric Aharonov-Bohm effect for  high velocity, and under  appropriate conditions that we provide in Theorem \ref{theor-3.1}. Our results settle the
 theoretical issues. It would be quite interesting if the existence of this fundamental phenomenon could be experimentally verified.

The results of this paper, as well as those of \cite{bw,bw2,bw3,w1},  are proven using the method to estimate the high-velocity
behaviour of solutions to the
Schr\"odinger equation and of the scattering operator that was introduced in
\cite{ew}, and was applied to time-dependent potentials in all space in \cite{wet}.

The paper is organized as follows. In Section 2 we state preliminary
results that we need. In Section 3 we obtain our estimates  for the
leading order at high velocity of the exact solution to the
Schr\"odinger equation and we use them to prove that quantum
mechanics rigorously predicts the existence of the electric
Aharonov-Bohm  effect under conditions that we provide. The main
result is Theorem \ref{theor-3.1} where we obtain our high-velocity
estimates for the exact solution to the Schr\"odinger equation that
give precise conditions for the validity of the Aharonov-Bohm
Ansatz, with an error bound in norm, given by ${\mathcal E}(v)$, that is uniform in time. In Theorems \ref{theor-3.2} and
\ref{theor-3.3} we obtain high velocity estimates for the wave and
the scattering operators that prove that these operators act as
multiplication by a constant phase given by  integrals in time of
the Aharonov-Bohm potential inside the tube, modulo an error that is uniform in time, and that
as before, is given ${\mathcal E}(v)$.

Finally some words about our notations and definitions. We denote by
$C$ any finite positive constant whose value is not specified. For
any $ x\in \ere^n, x \neq 0$, we denote, $\hat{x}:= x/|x|$.  For any
$ \v \in \ere^n$ we designate, $ v:= |\v|$.  As mentioned above, by $B_R$ we denote
the open ball of center $0$ and radius $R$. For any set $O$ we denote by  $\chi_O(x)$ the characteristic function of $O$ and by $F(x \in O)$ the operator of
multiplication by the characteristic function of $O$. By $\|\cdot
\|$ we denote the norm in $L^2(\Lambda)$ where, as above, $\Lambda:= \ere^n
\setminus K$. The norm of $L^2(\ere^n)$ is denoted by $\|\cdot
\|_{\ds L^2(\ere^n)}$.  For any open set, $O$, we denote by
$\mathcal H_s(O), \, s=1,2,\cdots$  the Sobolev spaces \cite{ad} and
by $\mathcal H_{s,0}(O)$ the closure of $C^\infty_0(O)$ in the norm
of $\mathcal H_{s}(O)$. By  $\mathcal{B}(O)$ we designate the Banach
space of all bounded operators on $L^2(O)$.  We denote by $\| \cdot \|_{\ds \mathcal{B}(\ere^n)}$ the
operator norm in $L^2(\ere^n)$.

We define the Fourier transform as a unitary operator on
$L^2(\ere^n)$ as follows,
$$
\hat {\phi}(p):= F \phi(p):= \frac{1}{(2 \pi)^{n/2}} \int_{\ere^n}
e^{-i p\cdot x} \phi (x)\, dx.
$$

We define functions of the operator $\mo:=-i \nabla$ by Fourier
transform,
$$
f(\mo) \phi:= F^\ast f(p) F \phi, \, D(f(\mo)):= \{ \phi \in
L^2(\ere^n): f(p) \,\hat{\phi}(p) \in L^2(\ere^n) \},
$$
for every measurable function $f$.

Let us mention some related rigorous results on the Aharonov-Bohm
effect. For further references see \cite{bw,bw2,bw3}, and
\cite{w1}. In \cite{hel}, a semi-classical analysis of the
Aharonov-Bohm effect in bound-states in two dimensions is given. The
papers \cite{rou}, \cite{ry1}, \cite{ya1}, and \cite{ya2} study the
scattering matrix for potentials of Aharonov-Bohm type in the whole
space.

\section{Preliminary Results}
\sss
 We consider a non-relativistic particle, like an electron,
that propagates outside a bounded metallic tube, $K$,  in $\ere^n, n \geq 2$, with its axis along the vertical direction. In the propagation
domain $\Lambda := \ere^n \setminus K$ there is a
 time-dependent electric potential as in (\ref{1.2}). To simplify the notation we multiply both sides of  Schr\"odinger's equation  (\ref{1.0})  by
$\frac{1}{\hbar}$ and we write it as follows

\beq \label{2.0}
i \frac{\partial} {\partial t} \phi= \frac{1}{2m} \mo^2
\phi+V\phi,
\ene
with $m:= M/ \hbar$ and 
$$
V:= \frac{q}{\hbar} \mathbf V=   \Vab(vt,x)+ V_0(t,x),
$$
where,
$$
  \Vab(vt,x):= \frac{q}{\hbar}\, {\mathbf V}_{AB}(vt,x) = v Q(v t,x),
  $$
  with
  $$
  Q(vt,x):= \frac{q}{\hbar}\, {\mathbf Q}(vt,x),
 $$
 and
 $$
 V_0(t,x):=  \frac{q}{\hbar}\, {\mathbf V}_0(t,x).
 $$

\subsection{The Tube  $K$}
For any $x=(x_1,x_2,\cdots,x_n)\in \ere^n$ we denote by  $
\overline{x}:= (x_1,x_2,\cdots,x_{n-1})$. Let $D_1, D_2 \subset
\ere^{n-1} $ be bounded open sets with $D_1 \subset D_2$ and let $L
>0$. The metallic tube, $K$, is the set \beq K:= \left\{x\in \ere^n:
\overline{x}\in \overline{D_2}\setminus D_1, |x_n|\leq L/2 \right\}.
\label{2.1} \ene For example, $K$ can be a cylindrical tube with
with $D_1$ and $D_2$ balls in $n\geq 4$  or discs in the case $n=3$.
The hole of the tube is the set, \beq K_0:= \left\{ x \in \Lambda :
\overline{x} \in D_1, |x_n| \leq L/2 \right\}. \label{2.2} \ene

\subsection{The Electric Potential}
 The electric potential $V(t,x)$ is a real -valued function defined on $\Lambda$.
In the following assumptions we summarize the conditions on $V(t,x)$ that we need.

We denote by $\Delta$ the self-adjoint realization of
the Laplacian in $L^2(\ere^n)$ with domain $\mathcal  H_2(\ere^n)$.
We say  that the operator of multiplication by  a real valued function $f$  defined in $\Lambda$ is $\Delta$- bounded with relative bound
zero if  the extension of $f$ to $\ere^n$ by zero
is $\Delta-$ bounded with relative bound zero \cite{ka}. Using a extension
operator from $\mathcal H_2(\Lambda)$ to $H_2(\ere^n)$ \cite{tr} we
prove that this is equivalent to require that $f$ is  relatively bounded from
$\mathcal H_2(\Lambda)$ into $L^2(\Lambda)$ with relative bound
zero.

We always assume that the electric potential $V(t,x)$ satisfies the following assumptions.
\beq
V(t,x):= \Vab(vt,x)+ V_0(t,x),
\label{2.3}
\ene
where the Aharonov-Bohm potential is given by,
\beq
\label{2.4}
 \Vab(z,x):= v \,Q(z,x)
\ene
with  $Q(z,x)=0$ for
$ x \in \Lambda \setminus K_0$ and for each fixed $x,\, Q(z,x)$ is continuously differentiable in $z$ and
\beq
\left|Q(z,x)\right|+ \left|\frac{\partial}{\partial z}Q(z,x)\right|\leq C,
\label{2.4b}
\ene
for some constant $C$. Furthermore,  for each $t \in \ere$  the operator of multiplication by  the function  $V_0(t,x)$ is $\Delta$-bounded
with relative bound
zero and the operator valued function
\beq
 t\rightarrow  V_0(t,x)\left( -\Delta+1\right)^{-1},
\label{2.5}
\ene
is  continuously differentiable in $ t \in \ere$, with values in ${\mathcal B}(\ere^n)$.
Moreover, there are $L_1 >L_0 >0$  such that $B_{L_1} \subset K_0$  and $ Q(z,x)=0, |z|\geq
L_0$. Furthermore,  for  $|z| <  L_0,\,\, Q(z,x)= Q_0(z)$   for $ x \in B_{L_1}$, where $Q_0(z)$ is a continuously differentiable  function
that vanishes for $|z| \geq L_0$. Note that $z=vt$
is the  distance along the classical trajectory of an electron that propagates with velocity $\v$.

Furthermore, we assume that,
 \beq
 \left\|V_{0}(t,x)\left( -\Delta+1\right)^{-1} F(|x| \geq r)\right\|_{\mathcal B(\ere^n)} \leq C (1+|t|)^{\mu} \, (1+r)^{- \rho}, \quad r \geq 0,
 \label{2.6}
 \ene
where $ \rho > 0, \mu \in \ere,$ and $  \rho- \mu >1$.

Remark that condition (\ref{2.6} is equivalent to the following assumption \cite{rs3}
\beq
 \left\|V_{0}(t,x)  F(|x| \geq r) \left( -\Delta+1\right)^{-1}\right\|_{\mathcal B(\ere^n)} \leq C (1+|t|)^{\mu} \, (1+r)^{- \rho}, \quad r \geq 0.
 \label{2.7}
 \ene
Condition (\ref{2.7}) has a clear intuitive meaning. It is an assumption on   the decay of $V_{0}$ at infinite.
However, in the proofs below we use the equivalent statement
(\ref{2.6}) because it is technically  more convenient.

Note that when $ \mu > 0$  the potential $V_{0}(t,x)$ can grow in time.
The physical reason for this is that, as in this case $V_{0}(t,x)$ goes to zero fast as $|x| \rightarrow \infty$,  in the 
high-velocity limit the
electron  leaves the interacting region, where $V_{0}(t,x)$ is strong, in a very small time, and then, the grow in time of $V_{0}(t,x)$ does not affects
the trajectory of the electron. When $ \mu < 0$,  $V_{0}(t,x)$ can go  to zero slowly  as $|x| \rightarrow \infty$, but this is compensated by the fact
that it goes to zero as time $ |t| \rightarrow \infty$. Note that along the classical trajectory, $ x= \v t,
V_{0}(t, \v t )$ decays as $1/t^{\rho- \mu}$ with
$\rho- \mu >1$, and hence, the effect of $V_{0}(t,x)$ is effectively of short-range, and, as we will see, the interacting evolution is  well approximated
by the free evolution, on spite of the fact that for each fixed time $V_{0}(t,x)$ can decay slowly as  $|x| \rightarrow \infty$.

The electron is inside the
tube during a time interval, around zero, of the order $1/v$. Hence,
$\Vab$ is different from zero only during a time interval of the
order $1/v$. Since as $v$ increases the time that $\Vab$ acts on the
electron decreases as $1/v$, in order that its effect does not
disappear for large $v$, it is necessary that the strength of $\Vab$
increases with the velocity $v$. Finally, note  that $V(t,x)$ depends on $v$ through $\Vab$. To simplify the notation we do not
make explicit this dependence on $v$.

Sufficient conditions for a multiplication by a function operator, $f$, to  be $\Delta$- bounded with relative bound zero are well known. For example
\cite{rs2}, for $n=3$, this is true if $f \in L^2(\ere^3)$ and for $ n \geq 4$ if $ f\in L^{p}(\ere^n)$ with $ p > n/2$. The function in (\ref{2.5}) is
 continuously differentiable, for example, if  $ t \rightarrow V_0(t,x)$ is a continuously differentiable function with values
 in $ L^2(\ere^3 $ for $ n=3$ and  in
  $L^{p}(\ere^n)$ with $ p > n/2$ for $ n\geq 4$. Obviously, we can replace  $L^2(\ere^3)$ by  $L^\infty(\ere^3)$ and $ L^{p}(\ere^n)$ by
   $L^{\infty}(\ere^n)$ in the conditions above, or by the sum of  potentials of this type. For more general sufficient conditions see \cite{sch}.

\subsection{The Unitary Propagator}
We define the unperturbed quadratic form,
$$
 h_0(\phi,\psi):= \frac{1}{2
m}(\mo \phi,\mo \psi), \, D(h_0):= \mathcal H_{1,0}(\Lambda).
$$
The associated positive operator in $L^2(\Lambda)$
\cite{ka}, \cite{rs2} is

$$
\frac{-1}{2m} \Delta_D,
$$
where $\Delta_D$ is  the  Laplacian with Dirichlet boundary condition on $\partial
\Lambda$. Note that the functions in  $\mathcal H_{1,0}(O)$ vanish
in trace sense in the boundary of $O$. By elliptic regularity \cite{ag},
$D(\Delta_D)= \mathcal H_2(\Lambda) \cap \mathcal H_{1,0}(\Lambda)$.

We define  the perturbed Hamiltonian as follows,
\beq
H(t):= \frac{-1}{2m} \Delta_D+ V(t,x),\quad t \in \ere,
\label{2.9}
\ene
with domain, $D(H(t)):= D(\Delta_D)= \mathcal H_2(\Lambda) \cap \mathcal H_{1,0}(\Lambda)$ independent of $t$.
Since $V(t,x)$ is $\Delta-$ bounded with relative bound zero it follows from Kato-Rellich's theorem \cite{ka,rs2} that $H(t)$ is self-adjoint and bounded below
in $L^2(\Lambda)$. Note that $H(t)$ is the physical perturbed Hamiltonian divided by $\hbar$. This is so, because we obtained equation (\ref{2.0})  multiplying both sides of the Schr\"odinger equation  (\ref{1.0}) by $\frac{1}{\hbar}$.  

We define the Hamiltonian $H(t)$  in $L^2(\Lambda)$  with Dirichlet boundary
condition at $\partial \Lambda$, i.e. $\phi=0$ for $x \in \partial
\Lambda$. This is the standard boundary condition that corresponds
to an impenetrable tube $K$. It implies that the probability that
the electron is at the boundary of the tube is zero.

It follows from  Theorem X.70 and from the proof of theorem X.71 of \cite{rs2} that under our conditions there exists a  unitary propagator
$U(t,q), t,q \in \ere$ such that:
\begin{enumerate}
\item  $ U(t,q), t, q \in \ere$ is a two-parameter family of  unitary operators on $L^2(\Lambda)$.
\item  $U(t,q)\, U(q,r)= U(t,r),U(t,t)=I, \quad \forall\, t,q,r \in \ere$.
\item $U(t,q)$ is jointly strongly continuous in $t,q \in \ere$.
\item
$U(t,q) D(\Delta_D) \subset  D(\Delta_D), \, \forall \,\, t,q \in \ere$ and $\forall \phi \in D(\Delta_D)$
$$
i \frac{\partial }{\partial t}\, U(t,q)\phi= H(t)\,U(t,q)\phi, \quad i \frac{\partial }{\partial q}\, U(t,q)\phi=- U(t,q)  \,H(q)  \phi,  \quad t,q \in \ere.
$$
\end{enumerate}
The unitary propagator gives us the unique solution to  Schr\"odinger's equation (\ref{2.0}) with initial data $\phi \in D(\Delta_D)  $ at $t=q$
 and with Dirichlet boundary condition at
$ \partial \Lambda$.

\subsection{Propagation Estimates}

The free Hamiltonian is the self-adjoint operator in $L^2(\ere^n)$,
\beq
H_0:= -\frac{1}{2m} \Delta
\label{2.13}
\ene
where $\Delta$ is the self-adjoint realization of the Laplacian with domain $\mathcal H_2(\ere^n)$. The solution to the free Schr\"odinger equation,

\beq \label{2.14}
 i \frac{\partial}{\partial t} \phi (t,x) = H_0
\phi(t,x), x \in  \ere^n, t \in \ere, \phi(0,x)= \phi_0 \in D(H_0),
\ene
is given by
\beq
\phi(t,x)= e^{-it H_0}\, \phi_0.
\label{2.14b}
\ene

It follows by Fourier transform that under translation in
configuration or momentum space generated, respectively, by $\mo$
and $x$ we obtain

\beq \label{2.10}
 e^{i \mo \cdot \v t} \, f(x) \, e^{-i \mo \cdot \v
t}= f(x+\v t), \ene
\beq \label{2.11}
 e^{-im\v\cdot x}\, f(\mo )\,
e^{im\v\cdot x}= f(\mo +m \v ),
\ene
and, in particular,
\beq
e^{-im\v\cdot x}\, e^{-it H_0 } \, e^{im\v\cdot x}= e^{-i mv^2
t/2}\, e^{-i\mo\cdot \v t}\,
 e^{-i tH_0}.
\label{2.12}
\ene

We need  the following lemma from \cite{w1}.
\begin{lemma}\label{lemm-2.0}

For any $f \in C^\infty_0(B_{m\eta})$ for some $\eta > 0$ and any $ j=1,2,\cdots $ there is a constant $C_j$ such that the following estimate holds
\beq
\left\| F(x \in \tilde{M})\, e^{-it H_0}\, f\left(\frac{\mo -m \tbv}{\tv}\right)\, F(x\in M )\right\|_{{\mathcal B}(\ere^n)}
 \leq C_j\, \left(1+ \lambda \tv + \eta \tv^2 |t|\right)^{-j},
\label{2.12b}
\ene
for any $\tbv \in \ere^n \setminus 0, \tv :=|\tbv|, t \in \ere$, and any measurable sets $\tilde{M}, M$ in $\ere^n$ such that $\lambda:=\hbox{\rm dist}\,\left(\tilde{M}, M+\tbv t \right)
-\eta \tv |t|\geq 0$.
\end{lemma}

\noindent{\it Proof:} This is the particular case of Lemma 2.1 of \cite{w1} with  $\rho=1 $ and $\tv=|\tbv|$. Note
that the proof in $n$ dimensions is the same as the one in two
dimensions given in \cite{w1}.

\begin{lemma} \label{lemm-2.1}

For any $f \in C^\infty_0(B_{m\eta})$ for some $0 <\eta < 1/8$, and for any $j=1,2,\cdots$
there is a constant $C_j$ such that
\beq
 \left\| F\left(|x- \tbv t| >
\frac{|\tv t|}{4}\right) e^{-i t H_0} \, f\left(\ds \frac{\mo-m
\tbv}{\tv}\right)\, F\left( |x| \leq
|\tv t|/8\right)\right\|_{{\mathcal B}(\ere^n)} \leq C_j (1+| \tv^2 t|)^{-j},
\label{2.16}
\ene for
 $\tv:=|\tbv| > 0$.
\end{lemma}

\noindent {\it Proof:} The lemma follows from Lemma \ref{lemm-2.0} with $\tilde{M}= \{|x-\tbv t|> |\tv t|/4  \}$ and $ M=\{|x|\leq |\tv t|/8 \}$.
Observe that $\lambda:=\hbox{\rm dist}\,\left(\tilde{M}, M+\tbv t \right)
-\eta \tv |t| \geq |\tv t|(1/8- \eta)$.

\bull

Recall that ${\mathcal E}(v)$ was defined in (\ref{3.11b}).

\begin{lemma} \label{lemm-2.2}

Let $f \in C^\infty_0(B_{m\eta}), 0 < \eta < 1/8 $. Suppose that $\mathcal V(t,x)$ satisfies (\ref{2.6}) or,
equivalently, (\ref{2.7}). Then, for any compact set $ D \subset
\ere^n$  and any $ \tv _0 > 0 $, there is a constant $C$ such that for all $\tv
\geq \tv_0 $,
\beq
 \int_{-\infty}^{\infty}\, dt\, \| \mathcal V(t,x) e^{ -i t H_0} \, e^{im \tbv\cdot x} \,
f\left(\ds \frac{\mo}{\tv}\right)  \phi \|_{L^2(\ere^n)} \leq C
 \, \|\phi\|_{ \mathcal H_2(\ere^n)}\, {\mathcal E}(\tv),
\label{2.17}
\ene
 for all     $\phi \in  \mathcal
H_2(\ere^n)$ with support in $ D$.

Furthermore , suppose that $\mathcal V(t,x)$ satisfies,
\beq
 \left\| \mathcal V(t,x)  F(|x| \geq r)\right\|_{\mathcal B(\ere^n)} \leq C (1+|t|)^{\mu} \, (1+r)^{- \rho}, \quad r \geq 0,
 \label{2.17b}
 \ene
 where $ \rho > 0,\, \mu \in \ere,$ and $  \rho- \mu >1.$
Then, for any compact set $ D \subset
\ere^n$  and any $ \tv _0 > 0 $, there is a constant $C$ such that for all $\tv \geq \tv_0 $,
\beq
 \int_{-\infty}^{\infty}\, dt\, \| \mathcal V(t,x) e^{ -i t H_0} \, e^{im \tbv\cdot x} \,
f\left(\ds \frac{\mo}{\tv}\right)  \phi \|_{L^2(\ere^n)} \leq C
 \, \|\phi\|_{ L^2(\ere^n)}\,{\mathcal E}(\tv),
\label{2.17c}
\ene
 for all     $\phi \in L^2(\ere^n)$ with support in $ D$.
\end{lemma}

\noindent {\it Proof:} It follows from (\ref{2.11}) that,
\beq
\begin{array}{c}
 \mathcal V(t,x) e^{ -i t H_0} \, e^{im \tbv\cdot x} \,
f\left(\ds \frac{\mo}{\tv}\right)  \phi =  e^{im \tbv\cdot x}  \mathcal V(t,x)\, (-\Delta+1)^{-1} e^{-it (\mo+ m \tbv)^2/2m } f\left(\ds
\frac{\mo}{\tv}\right)\, (-\Delta+1)\, \phi =\\\\
  e^{im \tbv\cdot x}  \mathcal V(t,x)\, (-\Delta+1)^{-1}\,e^{-im \tbv\cdot x}\,
  e^{-it H_0 }\,  f\left(\ds \frac{\mo-m
\tbv}{\tv}\right)\,   e^{im \tbv\cdot x} \,  (-\Delta+1)\, \phi.
\end{array}
\label{2.17d}
\ene
Then, we have that,
\beq
 \left\| \mathcal V(t,x) e^{ -i t H_0}\, e^{im\tbv\cdot x}
f\left(\ds \frac{\mo}{\tv}\right)  \phi \right\|_{L^2(\ere^n)}\leq
I_1+I_2+I_3, \label{2.18} \ene
where,
\beq I_1:=  \left\| \mathcal
V(t,x) (-\Delta+1)^{-1} F\left(|x- \tbv t| > \frac{|\tv
t|}{4}\right)  e^{ -i t H_0} f\left(\ds \frac{\mo-m
\tbv}{\tv}\right)\, F\left( |x| \leq |\tv
t|/8\right)\right\|_{\mathcal B(\ere^n)}\,  \|\phi \|_{\mathcal
H_2(\ere^n)},
\label{2.19}
\ene
\beq I_2:= C\,  \left\| \mathcal
V(t,x) (-\Delta+1)^{-1} \right\|_{\mathcal B(\ere^n)}\,   \| F\left(
|x| > |\tv t|/8\right) (-\Delta+1) \phi \|_{L^2(\ere^n)},
\label{2.20}
\ene
\beq I_3:= C  \left\| \mathcal V(t,x)
(-\Delta+1)^{-1} F\left(|x- \tbv t| \leq \frac{|\tv
t|}{4}\right)\right\|_{\mathcal B(\ere^n)}\,  \|\phi \|_{\mathcal
H_2(\ere^n)}. \label{2.21} \ene
By (\ref{2.6}) with $r=0$ and
(\ref{2.16}),
\beq I_1  \leq \, C_j \, (1+ \tv_0 |\tv t|)^{-j} \, \|\phi
\|_{\mathcal H_2(\ere^n)}, \quad j=1,2,\cdots. \label{2.22} \ene
Since   $\phi$ has compact support in $D$,
$$
\| F(|x|> |\tv t|/8) (-\Delta+1)\phi\|_{L^2(\ere^n)} \leq C_j
(1+|\tv t|)^{-j} \| (1+|x|)^j (\Delta+1)\phi \|_{L^2(\ere^n)} \leq
C_j\,(1+|\tv t|)^{-j} \, \|\phi\|_{ \mathcal H_2(\ere^n)}, \quad j=1,2,\cdots.
$$
Hence, by (\ref{2.6}) with $r=0$,
\beq
I_2 \leq \, C_j \, (1+|\tv t|)^{-j} \,  \|\phi \|_{\mathcal H_2(\ere^n)},    \quad j=1,2,\cdots.
\label{2.23}
\ene
It follows from (\ref{2.22}, \ref{2.23}) that
\beq
\int_{-\infty}^{\infty}\, dt \, (I_1+ I_2)\,\leq \, C \, \frac{1}{\tv}\,  \int_{-\infty}^{\infty}\,dz \,(1+|z|)^{-2}\,  \|\phi\|_{ \mathcal H_2(\ere^n)}
= C \frac{1}{\tv}\,  \|\phi\|_{ \mathcal H_2(\ere^n)}.
\label{2.24}
\ene

Furthermore, by (\ref{2.6})
\beq
\int_{-\infty}^{\infty}\, dt \, I_3 \leq C \, \int_{-\infty}^{\infty}\, dt \,(1+|t|)^{\mu} \, (1+|\tv t|)^{-\rho } \,
 \|\phi\|_{ \mathcal H_2(\ere^n)}\,\leq C \,   \|\phi \|_{\mathcal H_2(\ere^n)} \,
{\mathcal E}(\tv).
\label{2.25}
\ene
Equation (\ref{2.17}) follows from (\ref{2.18}, \ref{2.24}, \ref{2.25}). Finally, (\ref{2.17c}) is proven in the same way but, as in this case the
regularization $(-\Delta+1)^{-1}$ is not needed, we obtain the norm of $\varphi$ in $L^2(\ere^n)$.

\subsection{The Wave and Scattering Operators}

Let $J$ be the identification operator from $L^2(\ere^n)$ onto
$L^2(\Lambda)$ given by multiplication by the characteristic
function of $\Lambda$. The wave operators are defined as follows,

\beq
 W_{\pm}:= \hbox{\rm s-}\lim_{t \rightarrow \pm \infty} U(0,t)\, J\, e^{-it H_0},
\label{2.26}
\ene
provided that the strong limits exist. It follows from the Rellich local compactness theorem \cite{ad, rs3} that $J$ can be replaced by the
operator of multiplication by any function $\chi \in C^\infty(\ere^n)$ that satisfies  $\chi(x)=0$ in a bounded neighborhood of
$K$ and $\chi (x)=1$ for $ x$ in the complement of another bounded  neighborhood of $K$,

\beq
 W_{\pm}:= \hbox{\rm s-}\lim_{t \rightarrow \pm \infty} U(0,t)\, \chi  \, e^{-it H_0}.
\label{2.26b}
\ene

\begin{lemma}\label{lemm-2.3}

The wave operators $W_{\pm}$ exist, they are partially isometric with initial subspace $L^2(\ere^n)$ and they satisfy the intertwining relations,
\beq
 U(t,0) \, W_{\pm} = W_{\pm}\,e^{-it H_0}.
\label{2.27}
\ene
\end{lemma}

\noindent{\it Proof:} It is enough to prove the existence of the $W_{\pm}$ for all functions of the type,
$$
 e^{i m \tbv \cdot x}f\left(\frac{\mo}{\tv}\right)  \phi,
$$
with $\phi \in \mathcal H_2(\ere^n)$ of compact support and  $f \in C^\infty_0(B_{m\eta})$ where $\eta < 1/8$ because the set of all linear combinations of these 
functions is dense in $L^2(\ere^n)$.

By  equation (\ref{2.26b}) and Duhamel's formula,
\beq
W_{\pm}\,e^{i m \tbv \cdot x} f\left(\frac{\mo}{\tv}\right)  \phi=
 e^{i m \tbv \cdot x} f\left(\frac{\mo}{\tv}\right)  \phi +i \int_0^{\pm \infty}\, dt \,U(0,t)\,  \left[H(t)\, \chi(x)-
 \chi(x) \, H_0\right] \, e^{-itH_0}\,
e^{i m \tbv \cdot x} f\left(\frac{\mo}{\tv}\right) \phi.
\label{2.29}
\ene
Since,
$$
 \left[H(t)\, \chi(x)- \chi(x) \, H_0\right] = V(t,x)- \frac{1}{2m} (\Delta \chi(x))- i \frac{1}{m} (\nabla\chi(x)) \cdot \mo,
 $$
the integral in the right-hand side of (\ref{2.29}) is absolutely convergent by Lemma \ref{lemm-2.2}. The fact that the $W_{\pm}$ are
partially isometric with initial subspace $L^2(\ere^n)$ follows from Rellich's local compactness theorem \cite{ad,rs3}, and the intertwining relations (\ref{2.27})
are immediate from the definition of $W_{\pm}$.

\bull

The scattering operator
is defined as
\beq
S:= W_+^\ast\, W_-.
\label{2.30}
\ene

\section{High-Velocity Estimates}
\sss
\subsection{High-Velocity  Solutions to the Schr\"odinger Equation}
At the time of emission, i.e., as  $t \rightarrow -\infty$, the electron
wave packet is far away from $K$  and  it does not  interact with it.
Therefore, it can be parametrised with kinematical variables and it
can be assumed that it follows the free evolution (\ref{2.14b})
of an asymptotic state, $\varphi_\v$, with velocity $\v$,
\beq
\psi_{\v,0}:=
e^{-itH_0} \, \varphi_\v,
\label{3.1}
\ene
where
\beq \label{3.2}
\varphi_{\v}:= e^{i m \v\cdot x}\,\varphi, \quad
\varphi \in L^2(\ere^n).
 \ene
Note that in the momentum
representation $ e^{im \v \cdot x}$ is a translation operator by the
vector $ m \v
 $, what implies that in this  representation
the asymptotic state (\ref{3.2}) is centered at the classical
momentum $m \v $,
$$
\hat{\varphi}_{\v}(p)= \hat{\varphi}(p-m\v).
$$
The exact electron
wave packet, $\psi_{\v}(x,t)$, satisfies the interacting
Schr\"odinger equation  (\ref{2.0}) for all times and as $t
\rightarrow -\infty$ it has to approach the incoming wave packet,
i.e.,
$$
\lim_{t \rightarrow -\infty}  \left\| \psi_{\v}-J
\psi_{\v,0}\right\|=0.
$$
This means that we have to solve the interacting Schr\"odinger equation
(\ref{2.0}) with initial conditions at minus infinity. This is
accomplished by the  wave operator $ W_{-} $. In fact, we have that,
\beq
\psi_{\v}= U(t,0)\, W_-\, \varphi_{\v},
\label{3.3}
\ene
because, as $ U(t,0)$ is unitary,
\beq \label{3.4}
\lim_{t \rightarrow - \infty}\left\|U(t,0) \,W_{-}\,
\varphi_{\v}- J \, e^{-it H_0 } \varphi_{\v} \right\|=0.
\ene
We prove in the same way that
\beq
 U(t,0)\, W_+\, \varphi_{\v}
\label{3.5b}
\ene
is the unique solution to the Schr\"odinger equation such that

$$
\lim_{t \rightarrow  \infty}\left\|U(t,0) \,W_{+}\,
\varphi_{\v}- J \, e^{-it H_0 } \varphi_{\v} \right\|=0.
$$

In order to isolate the electric Aharonov-Bohm effect we need to separate the
effect of $K$ as a rigid body  from that of the electric potential inside the hole
$K_0$. For this purpose, we need asymptotic states that have negligible
interaction with $K$ for all times. This is possible for large enough  velocities.

For any $\v \neq 0$ we denote,
\beq
\Lambda_{\hv}:= \{x \in \Lambda:
x+\tau \hv \in \Lambda,\, \forall \tau \in \ere\}.
\label{3.6}
\ene
Let us consider  asymptotic states (\ref{3.2}) where  $ \varphi$
has compact support contained in $\Lambda_{\hv}$. For the discussion
below it is  better to parametrise the free evolution of
$\varphi_{\v}$ by the distance along the classical trajectory, $z=v t$, rather than by the time $t$.
It follows from (\ref{2.12}) that at distance $z$ the state is given by,
\beq
 e^{ -i \frac{z}{v}
H_0}\, \varphi_{\v}= e^{im\v\cdot x}\,   e^{-i\frac{mzv}{2}}\, e^{
-i \frac{z}{v} H_0}\, e^{-i\mo\cdot z \hv}  \varphi.
\label{3.7}
\ene
Observe that
$e^{-i\mo\cdot z \hv}$ is a translation in a straight line along the
classical free evolution,
\beq
\left( e^{-i\mo\cdot z \hv}
\varphi \right)(x)=  \varphi(x-z\hv).
\label{3.8}
\ene
The
term $ e^{ -i \frac{z}{v} H_0}$ gives raise to the
quantum-mechanical spreading of the wave packet. For high velocities
this  term is one order of magnitude smaller than the classical
translation, and if we neglect it we get that,
\beq
(e^{ -i
\frac{z}{v} H_0}\, \varphi_{\v})(x) \approx  e^{i\frac{mzv}{2}}\,
\varphi_{\v}(x-z\hv), \,\, \hbox{\rm for large}\,\, v.
\label{3.9}
\ene
We see that, in this approximation, for high velocities our
asymptotic state evolves along the classical trajectory, modulo the
global phase factor $e^{i\frac{mzv}{2}}$ that plays no
role. The key issue is that the support of our incoming wave packet
remains in $\Lambda_{\v}$ for all distances, or for all times, and
in consequence it has no interaction with $K$. We can expect that
for high velocities  the exact solution  $\psi_{\v}$ (\ref{3.3})
to the interacting Schr\"odinger equation (\ref{2.0})  is close to
the incoming wave packet $\psi_{\v,0}$ and that, in consequence, it
also has negligible interaction with $K$, provided, of course,  that
the support of $\varphi$ is contained in $\Lambda_{\v}$. Below we
give  rigorous ground for this heuristic picture proving that in
the leading order $\psi_{\v}$ is not influenced by $K$ and  that it
only contains information on the electric potential inside $K_0$.

\subsection{The Aharonov-Bohm Ansatz}
Aharonov and Bohm \cite{ab} observed that in a region of space where there is a potential $V(t)$ that is independent of $x$ the solution to the
Schr\"odinger equation  (\ref{2.0})  with $\phi(0)= \phi_0$ is given by,
$$
e^{-i \,\int_0^t \, ds \,V(s)} \, e^{-it H_0}\, \phi_0.
$$
We define,
 \beq
F_-(t):= v \int_{-\infty}^{t}\, Q_0(v s)\, ds.
\label{3.9b}
\ene
Note that,
\beq \begin{array}{c}
F_-(t)=0, \quad t \leq - L_0/v,\\\\
F_-(t)= F(L_0/v)= \Phi, \quad t \geq L_0/v,
\end{array}
\label{3.9c}
\ene
where,
\beq
\Phi:=\int_{-L_0}^{L_0}\, Q_0(z).
\label{3.9d}
\ene
 We define the following approximate solution to the Schr\"odinger equation (\ref{2.0}),
\beq
\psi_{AB,\v}(t,x) := e^{-i F_-(t)}\, e^{-it H_0}\, \varphi_{\v},
\label{3.9e}
\ene
where $\v$ is such that $ B_{L_1}\subset \Lambda_{\v}$. For example, we can take $\v$ along the vertical direction $x_n$ or slightly tilted with
respect to $x_n$. Furthermore, we assume that $\hbox{\rm support} \, \varphi \subset B_R$ for some $R < L_1-L_0$.
Suppose for the moment that $V_0=0$. For $t \leq -  L_0/v$, $\Vab=0$ and then,
(\ref{2.0}) is just the free Schr\"odinger equation (\ref{2.14}). But as for $ t\leq - L_0/v, \,F_-(t)=0,\, \psi_{AB,\v}$ is also a solution to the free Schr\"odinger
equation. Moreover, as $\hbox{\rm support} \, \varphi \subset B_R$, we have that according to the classical free evolution with velocity $\v$,
 for $|t| < L_0/v $ the electron is  inside the ball  $B_{R+L_0} \subset B_{L_1}$. But, since in $B_{L_1}, \,\Vab = v \, Q_0(v t)$ we can expect
 that $\psi_{AB,\v}$ is a good approximation to the exact solution for $|t| < L_0/v$. Finally, as for  $t \geq L_0 /v, \, \Vab=0$  we can expect that
 $$
 e^{-i(t- L_0/v) H_0} e^{-i \Phi} e^{-i (L_0/v) H_0}\varphi_{\v},
 $$
is a good approximation to the exact solution for $ t \geq L_0/v$. But,
$$
e^{-i(t- L_0/v) H_0} e^{-i \Phi} e^{-i (L_0/v) H_0}\varphi_{\v}= \psi_{AB,\v}, \quad \hbox{\rm for}\, t \geq L / v.
$$
Furthermore, as $V_0$ is uniformly bounded in $v$, we can expect that for high velocity it gives a contribution that does not appear in the leading
order of the solution. These considerations motivate the introduction of the following Aharonov-Bohm Ansatz.

\begin{Ansatz}\label{def-3.1}{\rm
Let $\v\in \ere^n \setminus 0$ be such that $B_{L_1}\subset \Lambda_\v$. Let $\varphi \in \mathcal H_2(\ere^n)$ satisfy $\hbox{\rm support}\,
\varphi \subset B_{R}$,
where $0 < R < L_1 -L_0$. Let $\psi_\v := U(t,0)\, W_- \,\varphi_\v$ be the solution to the Schr\"odinger equation that behaves like
$\psi_{\v,0}:= e^{-it H_0}\, \varphi_\v$ as  time goes to minus infinite. Then,
\beq
\psi_\v \approx \psi_{AB,\v}(t,x) := e^{-i F_-(t)}\, e^{-it H_0}\, \varphi_{\v},
\label{3.9f}
\ene
for large velocity, $v:=|\v|$, and uniformly in time.}
\end{Ansatz}

\subsection{Uniform Estimates for the Exact solution to the Schr\"odinger  Equation}
In this subsection we estimate the high-velocity solutions to the Schr\"odinger equation.

Let  $g \in C^\infty_0(\ere^n)$  satisfy, $ g(p)=1, |p|\leq m/32$ and $g(p)= 0, |p|\geq  \frac{m}{16} $. We denote,
\beq
\tilde{\varphi}:=
g(\mo / v) \, \varphi, \quad v>0.
\label{3.10}
\ene
By Fourier transform we prove that,
\beq
\left\| \tilde{\varphi}-\varphi
\right\|_{L^2(\ere^n)}\leq  \frac{C}{1+v^2}\,\|\varphi\|_{ \mathcal H_2(\ere^n)}.
\label{3.11}
\ene
We define,
\beq
F_+(t):= v \int_t^{\infty}\, Q_0(v s)\, ds.
\label{3.11a}
\ene
Note that,
\beq \begin{array}{c}
F_+(t)=0, \quad t \geq  L_0/v,\\\\
F_{+}(t)= F_+(-L_0/v)=  \Phi, \quad t \leq  -L_0/v,
\end{array}
\label{3.11ab}
\ene
where $\Phi$ is defined in (\ref{3.9d})

The next theorem is our main result where we give our high-velocity estimates, uniform in time,  for the exact solutions to the Schr\"odinger equation.
Recall that ${\mathcal E}(v)$ is defined in (\ref{3.11b}).
\begin{theorem}{Uniform Estimate of the Solutions.}\label{theor-3.1}

Let $ \v \in \ere^n \setminus 0$ be such that $ B_{L_1} \subset  \Lambda_\v$  and let $R$ satisfy, $ 0 < R <L_1 - L_0$. Then, there is a constant $C$
such that,
\beq \left\|U(t,0)\, W_{\pm}
\,\varphi_{\v} - e^{\pm i F_{\pm}(t)} \,  e^{-i t\,H_0} \varphi_\v \right\| \leq \,C \, \|\varphi\|_{ \mathcal H_2(\ere^n)}\, \,{\mathcal E}(v),
\label{3.17}
\ene
for all $\varphi \in
{\mathcal H}_2(\ere^n)$ with support contained in $ B_R$.
\end{theorem}

\noindent {\it Proof:} By (\ref{3.10}, \ref{3.11}) it is enough to prove the theorem for $ e^{im \v\cdot x}\, \tvf$. Let $\chi \in C ^\infty(\ere^n)$  satisfy  $\chi(x)=0$ in a bounded neighborhood of
$K$ and $\chi (x)=1$ for $ x \in \left\{ x: x= y +\hv \tau, y \in \, \overline{B_R} , \tau \in \ere \right\} \cup \left\{x: |x| \geq N \right\}$ with
$N$ so large that $K \subset B_N$.

By equation (\ref{2.26b})  and  Duhamel's formula,
\beq\begin{array}{c}
U(t,0)\, W_{\pm}
\, e^{im \v\cdot x}\,\tvf - \chi(x) e^{\pm iF_{\pm}(t)}\,  e^{-i t\,H_0} \, e^{im \v\cdot x}\,\tvf=
i \int_0^{\pm \infty}\,dr\, U(t,t+r )\, \left( H(t+r)\chi - \chi H_0-\chi vQ_0(v(t+r)) \right)\,\\\\  e^{\pm i F_{\pm}(t+r)}\,\,  e^{-i(t+r) H_0}\,
\, e^{im \v\cdot x}\,\tvf.
\end{array}
\label{3.18}
\ene
Furthermore,
\beq
U(t,0)\, W_{\pm}
\,e^{im \v\cdot x}\,\tvf  - \chi(x) \, e^{\pm i F_{\pm}(t)}\,\,  e^{-i t\,H_0} e^{im \v\cdot x}\,\tvf =  i \int_0^{\pm \infty}\,dr \, U(t,t+r )\, (T_1 +T_2 + T_3) \,
\label{3.19}
\ene
where,
\beq
\begin{array}{c}
T_1:= \left(V_{0}(t+r,x) \chi(x) - \frac{1}{2m}(\Delta \chi)(x) \right)\,  e^{\pm i F_{\pm}(t+r)}\,\, e^{-i(t+r) H_0} \,
e^{im \v\cdot x}\,\tvf - \\\\
\frac{i}{m}\,
 (\nabla \chi)(x)\cdot \,  e^{\pm i F_{\pm}(t+r)}\,\, e^{-i(t+r) H_0}\, \, e^{im \v\cdot x}\, \mo \,\tvf,
\end{array}
 \label{3.20}
 \ene
 \beq
 T_2:= -i (\nabla \chi)(x)\cdot \v \,  e^{\pm i F_{\pm}(t+r)}\,\, e^{-i(t+r) H_0} \, e^{im \v\cdot x}\,\tvf,
 \label{3.21}
 \ene
\beq
 T_3:=  \chi_{(-L_0 /v, L_0/v)}(t+r)\,(\Vab(t+r,x)- v Q_0(v(t+r)))\chi(x) \,  e^{\pm i F_{\pm}(t+r)}\,\, e^{-i(t+r) H_0} \, e^{im \v\cdot x}\,\tvf.
 \label{3.21b}
 \ene
 By Lemma \ref{lemm-2.2} and as $U(t,q)$ is unitary, for $ v \geq 1$
 \beq
 \int_{-\infty}^\infty \ \, dt \,\left\|  U(t,t+r ) T_1 \right\| \leq \,C \, \|\varphi\|_{ \mathcal H_2(\ere^n)}\, {\mathcal E}(v).
\label{3.22}\ene
Moreover, as in the proof of equation  (2.65) of \cite{w1} we prove that for $v \geq 1$,
\beq
\int_{-\infty}^\infty \ \, dt \,\left\|  U(t,t+r ) T_2 \right\|=  \int_{-\infty}^\infty \ \, dt \,\left\| T_2 \right\| \leq \frac{C}{v} \,
\|\varphi\|_{ \mathcal H_2(\ere^n)}.
\label{3.23}
\ene
We give below the proof of this estimate, for the reader's convenience.

We define,
\beq
a(x):=|\nabla \chi(x)|.
\label{3.23-1}
\ene
Then,

\beq \label{3.23-2}
\int_{-\infty}^\infty \ \, dt \,\left\| T_2 \right\|  \leq \int_{-\infty}^\infty \ \, d\tau \, \left\| a(x)e^{ -i \frac{\tau}{v}\, H_0} \, e^{im \v\cdot x}\,\tvf
\right\|.
\ene

Arguing as in the proof of Lemma \ref{lemm-2.2}, but without introducing the regularization  $(-\Delta+1)^{-1}$ since $a(x)$ is bounded,
we prove that, 
\beq
\left\| a(x) \, e^{ -i \frac{\tau}{v}\, H_0} \, e^{im \v\cdot x}\,\tvf
  \right\| \leq C_l \, 
(1+|\tau| )^{-l} \,  \|\varphi\| , l=1,2,\cdots,
\label{3.23-3}
\ene
where we also used that $a(x)$ has compact support.
Moreover, as
$\chi (x)=1$ for  $ x \in \left\{ x: x= y +\hv \tau, y \in \, \overline{B_R}, \tau \in \ere \right\}$, we have that,
 $a(x+\hv \tau )\, \varphi(x)=0$. Hence, by (\ref{2.10}, \ref{2.12})
 
$$
 a(x) \, e^{ -i \frac{\tau}{v}\, H_0} \, e^{im \v\cdot x}\,\tvf =  a(x) \, e^{ -i \frac{\tau}{v}\, H_0} \, e^{im \v\cdot x}\,(\tvf
- \varphi)+ e^{im \v\cdot x}    e^{-i( \mo \cdot \hv \tau + m v \tau / 2 )}\, 
a(x+ \hv \tau )   \left(e^{-i H_0 \tau /v}-I \right)\, \varphi. 
$$
Then,
\beq
\left\|  a(x) \, e^{ -i \frac{\tau}{v}\, H_0} \, e^{im \v\cdot x}\,\tvf    \right\| \leq C \frac{(1+|\tau|)}{v}\,
 \|\varphi\|_{ \mathcal H_2(\ere^n)},
\label{3.23-4}
\ene
where we used (\ref{3.11}). By (\ref{3.23-3}) and (\ref{3.23-4}),
\beq
\left\| a(x) \,e^{ -i \frac{\tau}{v}\, H_0} \, e^{im \v\cdot x}\,\tvf   \right\| \leq C_{\delta ,l} \,\frac{1}{v^{\delta}}
(1+|\tau| )^{-l}\, \|\varphi\|_{ \mathcal H_2(\ere^n)}, l=1,2,\cdots , 0 \leq \delta < 1.
\label{3.23-5}
\ene
We define,
\beq
 I(\v):=\int \, \gamma (\v, \tau ) \, d \tau,
\label{3.23-6}
\ene
where,
\beq
\gamma (\v ,\tau ):=\left[ \left\| a(x) \, e^{ -i \frac{\tau}{v}\, H_0} \, e^{im \v\cdot x}\,\tvf \right\|^2 +  v^{-4} (1+|\tau |)^{-4}\right]^{1/2}.
\label{3.23-7}
\ene
Equation  (\ref{3.23-5}) implies that,  $I(\v) \ < \infty$ and that
$ \lim_{ v \rightarrow  \infty} I(\v)= 0$. By  (\ref{2.10}, \ref{2.12})  we have that,
\beq
\left\| a(x) e^{ -i \frac{\tau}{v}\, H_0} \, e^{im \v\cdot x}\, H_0 \tvf  \right\| = \left\| a(x+ \hv \tau ) \, e^{ -i \frac{\tau}{v}\, H_0} \,  H_0\tvf \right\|.
\label{3.23-8}
\ene
 Hence,
\beq
\left|\frac{\partial}{\partial v} \gamma(\v ,\tau)\right|
\leq C \, \left[ \frac{|\tau|}{v^2}  \left\| a( x + \hv \tau )\, e^{-i  \frac{\tau}{v}  H_0 }
H_0 \,\tvf \right\|+ v^{-3}\, (1+|\tau|)^{-2} \right].
\label{3.23-9}
\ene
As in the proof of (\ref{3.23-3}) we prove that,
\beq
\left\| a(x) e^{ -i \frac{\tau}{v}\, H_0} \, e^{im \v\cdot x}\, H_0\, \tvf  \right\|   \leq C_l \, 
(1+|\tau| )^{-l} \,  \|\varphi\||_{ \mathcal H_2(\ere^n)} , l=1,2,\cdots.
\label{3.23-10}
\ene

By (\ref{3.23-8}-\ref{3.23-10}) we have that,
\beq
\left|\frac{\partial}{\partial v} \gamma(\v ,\tau)\right|
 \leq C\, v^{-2}\,
(1+|\tau |)^{-2}, v \geq 1,
\label{3.23-11}
\ene
and it follows that
\beq
\left|\frac{\partial}{\partial v} I(\v)\right| \leq C \, v^{-2}
\label{3.23-12}.
\ene
Hence,
\beq
I(\v)= \left| \int_{v}^{\infty} \frac{\partial}{\partial s} I(s \hv )
 \, ds \right| \leq C v^{-1}.
\label{3.23-13}
\ene
The estimate (\ref{3.23})  follows from  (\ref{3.23-2}, \ref{3.23-6}, \ref{3.23-7}) and (\ref{3.23-13}).

We have that,
\beq\begin{array}{c}
T_3= \chi_{(-L_0 /v, L_0/v)}(t+r)\, \chi_{\widetilde{B_{L_1}}}(x)  (\Vab(t+r,x)- v Q_0(v(t+r)))\chi(x) \\\\  e^{\pm i F_{\pm}(t+r)}\,\,
 e^{-i(t+r) H_0}\,g\left( \frac{\mo -m\v}{v}
\right)\,
  \chi_{B_R}(x) \, e^{im \v\cdot x}\,\varphi,
\end{array}
 \label{3.23b}
 \ene
where $\widetilde{B_{L_1}}$ is the complement of $B_{L_1}$. We take $g$ in (\ref{3.10}) with support in $B_{m\eta}$ with $ \eta \leq \hbox{\rm min}\,
[1/16, \frac{ L_1-L_0-R}{L_0}]$. We take in Lemma \ref{lemm-2.0} $ \tilde{M}= \widetilde{B_{L_1}}, M=B_R$ and $\tbv = \v$. Note that
for $|t+r|\leq L_0/v$, $\hbox{dist}\,\left( \widetilde{B_{L_1}}, B_R+\v (t+r)\right) -\eta v|t+r|\geq L_1-L_0-R- \eta L_0>0$. Then, by Lemma \ref{lemm-2.0},
$$
\|T_3\|\leq  C_j  v  \chi_{(-L_0 /v, L_0/v)}(t+r) \,(1+ v)^{-j} \,  \|\varphi\|_{ L^2(\ere^n)}, \quad j=1,2,\cdots,
$$
and then,
\beq
\int_{-\infty}^\infty \ \, dt \,\left\|  U(t,t+r ) T_3 \right\|\leq C_j \frac{1}{(1+v)^j}  \int_{-L_0}^{L_0}\, dz \,\|\varphi\|_{ L^2(\ere^n)}
 \leq \frac{C_j}{(1+v)^j} \, \|\varphi\|_{ L^2(\ere^n)}, \quad j=1,2,\cdots.
\label{3.23c}
\ene
 Let us denote by $\mathcal S$ the support of $1-\chi(x)$. Note that there is a $R_1 < R$  such that, $\hbox{support}\, \varphi \subset B_{R_{1}}$. Then,
\beq
\left\|(1-\chi(x))\,  e^{\pm i F_{\pm}(t)}\,\,  e^{-i t\,H_0} e^{im \v\cdot x}\,\tvf \right\|_{L^2(\ere^n)} \leq C \left\|\chi_{\mathcal S}(x)\,  e^{-i t\,H_0}
 g\left(\frac{\mo-m\v}{v}\right) \chi_{B_{R_1}}(x) \right\|_{{\mathcal B}(\ere^n)} \, \|\varphi\|_{ L^2(\ere^n)}.
 \label{3.23d}
\ene
Observe  that $\hbox{\rm dist}\, (\mathcal S, B_{R_1}+\v t)\geq R- R_1$, and that for $ |\v t| \geq 4 N$, $\hbox{\rm dist}\, (\mathcal S, B_{R_1}+\v t)
\frac{1}{2} |\v t|+ N-R_1 > \frac{1}{2} |\v t| $. Then, we can always take  $g$ with support in $B_{m\eta}$ with  $\eta$ so small that
$\hbox{\rm dist}\, (\mathcal S, B_R+\v t)- \eta |\v t|\geq \tilde{\rho} >0, \forall \v t $. Hence, by Lemma \ref{lemm-2.0} with $\tilde{M}=
{\mathcal S}, M=B_{R_1} $ and $\tbv=\v$ we have that

 \beq
 \left\|(1-\chi(x))\,  e^{\pm i F_{\pm}(t)}\,\,  e^{-i t\,H_0} e^{im \v\cdot x}\,\tvf   \right\|_{L^2(\ere^n)}
 \leq \, \frac{C_j}{v^j}\,
 \|\varphi\|_{L^2(\ere^n)} \quad j=1,2,\cdots.
\label{3.24}
\ene
Equation (\ref{3.17}) follows from (\ref{3.11}, \ref{3.19}, \ref{3.22}, \ref{3.23}, \ref{3.23c})  and (\ref{3.24}).
\subsection{High-Velocity Estimates of the Wave and the Scattering Operators}
Theorem \ref{theor-3.1} implies the following high-velocity estimates for the wave and the scattering operators.
\begin{theorem} \label{theor-3.2}

Let $ \v \in \ere^n \setminus 0$ be such that $ B_{L_1} \subset  \Lambda_\v$  and let $R$ satisfy, $ 0 < R < L_1 - L_0$. Then, there is a constant $C$
such that,
\beq \left\| e^{-im\v\cdot x}\, W_{\pm}
\,  e^{im\v\cdot x} \,\varphi - e^{\pm i F_{\pm}(0)} \, \varphi \right\| \leq \,C \, \|\varphi\|_{ \mathcal H_2(\ere^n)}\, \,{\mathcal E}(v),
\label{3.25}
\ene
\beq
 \left\| e^{-im\v\cdot x}\, W_{\pm}^\ast
\,  e^{im\v\cdot x} \,\varphi - e^{\mp i F_{\pm}(0)} \, \varphi \right\|_{L^2(\ere^n)} \leq \,C \, \|\varphi\|_{ \mathcal H_2(\ere^n)}\, \,{\mathcal E}(v),
\label{3.26}
\ene
for all $\varphi \in
{\mathcal H}_2(\ere^n)$ with support contained in $ B_R$.
\end{theorem}

\noindent{\it Proof:} Equations (\ref{3.25}) are just (\ref{3.17}) with $t=0$. to prove (\ref{3.26}) we denote,
$$
W_{\pm,\v}:= e^{-im\v\cdot x}\, W_{\pm}
\,  e^{im\v\cdot x}.
$$
Since the wave operators are partially isometric, $W_{\pm,\v}^\ast \, W_{\pm, \v}= I$. Then,
$$
\begin{array}{c}
\left\| \, W_{\pm,\v}^\ast
\,\varphi - e^{\mp i F_{\pm}(0)} \, \varphi \right\|_{L^2(\ere^n)}= \left\| \, W_{\pm,\v}^\ast
\,\varphi -  W_{\pm,\v}^\ast\, W_{\pm,\v}\,  e^{\mp i F_{\pm}(0)} \, \varphi \right\|_{L^2(\ere^n)} \leq \\\\
\left\| \left(\, e^{\pm i F_{\pm}(0)} - W_{\pm,\v} \right)\,  e^{\mp i F_{\pm}(0)} \, \varphi \right\|_{L^2(\ere^n)}\leq \,C \,
 \left\|\varphi\right\|_{ \mathcal H_2(\ere^n)}\, \,{\mathcal E}(v).
\end{array}
$$

\bull

Note that (see (\ref{3.9b}, \ref{3.9d}) and (\ref{3.11a})),

$$
\Phi = F_+(0)+F_-(0).
$$

\begin{theorem}\label{theor-3.3}

Let $ \v \in \ere^n \setminus 0$ be such that $ B_{L_1} \subset  \Lambda_\v$  and let $R$ satisfy, $ 0 < R < L_1 - L_0$. Then, there is a constant $C$
such that,
\beq \left\| e^{-im\v\cdot x}\,S
\,  e^{im\v\cdot x} \,\varphi - e^{- i \Phi} \, \varphi \right\| \leq \,C \, \|\varphi\|_{ \mathcal H_2(\ere^n)}\, \,{\mathcal E}(v),
\label{3.27}
\ene
for all $\varphi \in
{\mathcal H}_2(\ere^n)$ with support contained in $ B_R$.
\end{theorem}

\noindent{\it Proof:} The theorem follows from Theorem \ref{theor-3.2} and the following argument.
$$
\begin{array}{c}
\left\| e^{-im\v\cdot x}\, S
\,  e^{im\v\cdot x} \,\varphi - e^{-i  \Phi} \, \varphi \right\|_{L^2(\ere^n )}= \left\| W_{+,\v}^\ast \, W_{-, \v}   \,\varphi -
W_{+,\v}^\ast \, W_{+, \v} \, e^{-i  \Phi} \, \varphi \right\|_{L^2(\ere^n )} \leq \\\\
\left\| \left( W_{-, \v}-e^{-i F_-(0)}\right) \,\varphi -
\left( W_{+, \v}- e^{i F_+(0)}\right)\, \, e^{-i  \Phi} \, \varphi \right\|_{L^2(\ere^n )}\leq  \,C \, \|\varphi\|_{ \mathcal H_2(\ere^n)}\, \,{\mathcal E}(v).
\end{array}
$$

\begin{figure}
\setlength{\unitlength}{1cm}
\hspace{4cm}
\includegraphics[width=20cm,totalheight=20cm]{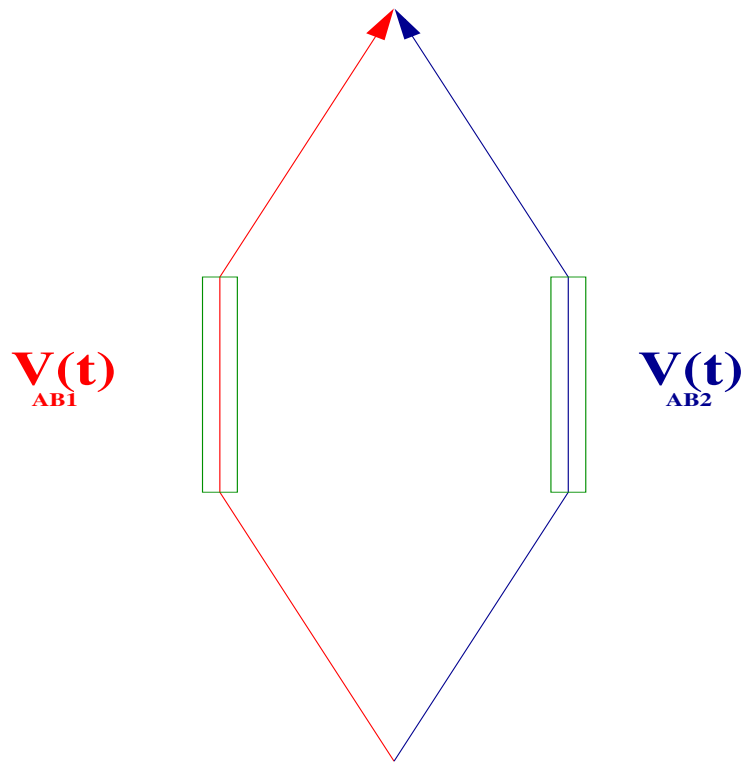}
\caption{The Electric Aharonov-Bohm Effect. Color Online.}
\end{figure}

\end{document}